\documentclass[preprint]{emulateapj}
\usepackage{epsfig,natbib}
\usepackage{amsmath,amsfonts,amssymb}
\usepackage{mathptmx}
\newcommand{\Msun}{\>{\rm M_{\odot}}}

\newcounter{species}
\def\ion#1#2{\setcounter{species}{#2}#1$\;${\sc\roman{species}}\relax}

\begin{document}

\title{SDSS~J092712.65+294344.0: Recoiling Black Hole or \\ A Sub-parsec Binary Candidate?}
\author{Tamara Bogdanovi\'c\altaffilmark{1}, Michael Eracleous\altaffilmark{2}, and Steinn Sigurdsson\altaffilmark{2}}

\altaffiltext{1}{Department of Astronomy, University of Maryland,
College Park, MD 20742-2421, e-mail: {\tt tamarab@astro.umd.edu}}

\altaffiltext{2}{Department of Astronomy and Astrophysics and Center
for Gravitational Wave Physics, Pennsylvania State University,
University Park, PA 16802, e-mail: {\tt mce, steinn@astro.psu.edu}}

\begin{abstract}

A search for recoiling supermassive black hole candidates recently
yielded the best candidate thus far, SDSS~J092712.65+294344.0 reported
by Komossa et al. Here we propose the alternative hypothesis that this
object is a supermassive black hole binary.  From the velocity shift
imprinted in the emission-line spectrum we infer an orbital period of
$\sim 190$ years for a binary mass ratio of 0.1, a secondary black
hole mass of $10^8~{\rm M}_\odot$, and assuming inclination and
orbital phase angles of $45^\circ$.  In this model the origin of the
blueshifted narrow emission lines is naturally explained in the
context of an accretion flow within the inner rim of the circumbinary
disk. We attribute the blueshifted broad emission lines to gas
associated with a disk around the accreting secondary black hole. We
show that, within the uncertainties, this binary system can be long
lived and thus, is not observed in a special moment in time. The
orbital motion of the binary can potentially be observed with the
$VLBA$ if at least the secondary black hole is a radio emitter. In
addition, for the parameters quoted above, the orbital motion will
result in a $\sim 100~{\rm km\,s^{-1}}$ velocity shift of the emission
lines on a time scale of about a year, providing a direct
observational test for the binary hypothesis.

\end{abstract}

\keywords{black hole physics -- galaxies: nuclei -- galaxies: 
individual (SDSS~J092712.65+294344.0) -- quasars: emission lines}

\section{Introduction}

The ``mass loss'' and gravitational rocket effect in the aftermath of
the coalescence of a supermassive black hole binary (SBHB) can have
profound effects on the properties of the nucleus of the host galaxy
and give rise to unique observational signatures \citep{rr89, milos05,
loeb07, sb08, lippai08, sk08, kl08, moh08, devecchi08,bl08,gm08}.

Since the emission of gravitational waves in the last stages of the
merger is not symmetric in general, the product of the merger can
receive a significant recoil velocity, up to a few $\times \,100~{\rm
  km\,s^{-1}}$ for low black hole spins or spin axes aligned with the
orbital axis \citep{herrmann07a, herrmann07b, baker06, baker07,
  gonzalez07b, koppitz07, rezzolla08}. In the special case of
maximally spinning, equal mass supermassive black holes (SBHs) with
their spin vectors in the orbital plane and directed opposite to each
other, the recoil speed can be up to $\sim$ 4000~ ${\rm km\,s^{-1}}$
\citep{campanelli07b} . The fraction of coalescences expected to
produce a remnant recoiling at $V>$ 1000~${\rm km\,s^{-1}}$ is about
$10\, \%$, assuming black hole spins of $cJ/GM^2 = 0.9$, mass ratio
range of 0.1 $<$ q $<$ 1, and arbitrary spin orientations
\citep{sb07,campanelli07a,baker08}.

Since the escape speed from most galaxies is less than 2000~${\rm
km\,s^{-1}}$ \citep{merritt04}, if high velocity recoils are common,
there should be many ``empty nest" galaxies, without a central
SBH. This is in contrast with the observation that almost all galaxies
with bulges have a central SBH \citep[for summary see][]{ff05}.
However, if following a galactic merger the two SBHs find themselves
in a gas rich environment, gas accretion torques will act to align the
spin axes with the orbital axis and thus, reduce the post-merger kick
to a value well below the galactic escape speed. This effect is
expected to increase the chance of retention of recoiling SBHs by
their host galaxies \citep{bogdanovic07}. Gas-poor mergers, on the
other hand can lead to large recoil speeds that launch the final black
hole out of the potential well of the host bulge or host galaxy.
 
\citet{bonning07} searched the Sloan Digital Sky Survey (SDSS) archive
for merger products with large velocity offsets from their host
galaxy. Among nearly 2600 objects, they found no convincing evidence
for recoiling SBHs in quasars, placing an upper limit of 0.2~$\%$ on
the probability of kicks with projected speeds greater than 800~${\rm
  km\,s^{-1}}$ and 0.04~$\%$ on the probability of kicks with
projected speeds greater than 2500~${\rm km\,s^{-1}}$. More recently,
Komossa et al. (2008), hereafter \citet{komossa08}, reported the
discovery of one such candidate, a SBH receding from its host galaxy
at a high projected speed of 2650 ${\rm km\,s^{-1}}$. This detection
has very important implications for the cosmological evolution of
SBHs. However, it is also remarkable given the likelihood of high
velocity kicks and a narrow observational window (in cosmological
terms) associated with this class of objects \citep{bl08}.  Combined
with several observational curiosities, they call the recoil idea into
question (see discussion in \S\,3). Here we propose that the peculiar
spectroscopic properties of SDSS~J092712.65+294344.0 \citep{am07} can
be explained in the context of a SBHB model. In \S\,\ref{S_model} we
describe the proposed model, in \S\,\ref{S_conclude} we discuss its
physical basis, revisit the recoiling black hole model, and close by
considering observational tests.

\section{A Supermassive Binary Black Hole Model for 
SDSS~J092712.65+294344.0}\label{S_model}

SDSS~J092712.65+294344.0 (hereafter J0927) exhibits an unusual optical
emission-line spectrum, which features two sets of lines offset by
2650~${\rm km\,s^{-1}}$ relative to each other. The ``redward'' system
consists of only narrow emission lines (r-NELs) with a full width at
half maximum (FWHM) of about 170~${\rm km\,s^{-1}}$ at a redshift of
$z=0.713$; this was identified by \citet{komossa08}, as the redshift
of the galaxy that hosted the recoiling SBH at its birth.  The
``blueward'' emission-line system comprises broad Balmer lines
(b-BELs) and narrow, high-ionization forbidden lines (b-NELs) at a
redshift of $z=0.698$ \citep[see Table~1 in][for the list of
lines]{komossa08}. In addition, the emission-line ratios of both the
blueward and redward system are consistent with photoionization by a
hard ionizing continuum typical of an active galactic nucleus (AGN).

According to the recoiling SBH interpretation of \citet{komossa08},
the b-BELs originate in the broad-line region (BLR) retained by the
SBH. Because of their lower FWHM in comparison to b-BELs, the b-NELs
were attributed to gas that is only marginally bound to the SBH. The
FWHM of the b-NELs was explained in the context of an accretion disk
that is expanding in the process of outward transport of angular
momentum, the associated outflows, and the swept up ISM. The r-NEL
lines were attributed to the narrow-line region that remains bound to
the host galaxy. The accreting SBH provides a source of ionizing
radiation for the b-BELs and b-NELs, as well as for the r-NELs, albeit
from a larger distance.

We find that, in the time that it takes the disk to expand to the
extent needed to explain the width of the b-NELs, the AGN has traveled
tens of kiloparsecs away from the host galaxy and thus, is not likely
to power the r-NELs. Combined with the narrow observational window and
specific configuration of the pre-coalescence binary, this presents a
challenge for the recoiling SBH model (see \S\,\ref{S_revisit} for
more detailed discussion). Note however that the outflow origin of the
b-NELs cannot be excluded based on the available data.

We make the alternative suggestion that the observed velocity shift
represents the projected orbital velocity of a bound black hole
pair. In this model, the emission lines of the blueward system (b-NELs
and b-BELs) originate in gas associated with the less massive,
secondary black hole, while the r-NELs originate in the ISM of the
host galaxy (Figure~\ref{fig_sketch}).  The accreting secondary SBH is
the main source of ionizing radiation while the primary SBH is either
quiescent or much fainter than the secondary (see the discussion in
\S\,3). From the observed X-ray luminosity of J0927 \citep[$L_{\rm X}
= 5 \times 10^{44}\;{\rm erg\,s^{-1}}$;][]{komossa08} we infer the
bolometric luminosity of the secondary\footnote{Using the average
quasar spectral energy distributions of \citet{elvis94}, we find
that that the bolometric luminosity is related to the X-ray
luminosity via $L_{\rm bol}\approx 14\; L_{\rm X}$. Here, the X-ray
luminosity is measured in the 0.1--2.4~keV band observable by {\it
ROSAT}.}  and derive a lower limit on its mass of $M_2 \gtrsim
5\times 10^7 \Msun$, based on the Eddington limit. We interpret the
observed velocity separation of the two emission line systems as the
projected velocity of the secondary relative to the center of
mass. Assuming a circular orbit, the projected velocity of the
secondary, $u_2$, is related to its orbital velocity, $V_2$, via
$u_2=V_2 \,\sin{i}\,\sin{\phi} = 2650\,{\rm km\,s^{-1}}$, where $i$ is
the inclination of the orbital axis of the binary relative to the line
of sight, and $\phi$ is the orbital phase at the time of the
observation ($\phi=0$ corresponds to conjunction). We derive the
following expressions for the binary separation and orbital period as
\begin{equation}
a\approx 0.16 \; {M_{2,8}\over 1+q} \left({0.1\over q}\right)\left({\sin i \over \sin 45^\circ}\;{\sin \phi\over \sin 45^\circ}\right)^2\; {\rm pc}\; ,
\label{eq1}
\end{equation}
\begin{equation}
P  \approx  190 \; {M_{2,8}\over (1+q)^2} \left({0.1\over q}\right)\left({\sin i\over \sin 45^\circ}\;{\sin \phi\over \sin 45^\circ}\right)^3\;{\rm yr}\, .
\label{eq2}
\end{equation}
In the above expressions, $q \equiv M_2 / M_1 \leq 1$ is the binary
mass ratio, and $M_{2,8}$ is the mass of the secondary SBH in units of
$10^8~{\rm M}_\odot$. We refrain from using the AGN scaling relations
based on the H$\beta$ line width and the continuum luminosity
\citep{kaspi05} to estimate the size of the BLR. A scenario in which
the size of the BLR is determined by the SBHB dynamics most likely
does not represent a ``typical'' AGN and hence, these or similar
relations may not hold. For example, the BLR around the secondary SBH
can be truncated by the tidal forces from the primary. This would
result in broader lines, compared to a typical AGN with a black hole
of the same mass.

If the measured FWHM of the [\ion{O}{3}]$\;\lambda$5007 lines is used
as an indicator of the stellar velocity dispersion \citep{nelson00},
the $M-\sigma$ relation \citep[][for example]{tremaine02} implies a
mass for the primary SBH that is $\sim 3$ orders of magnitude below
what we adopt in our model. However, such a discrepancy is not
unprecedented; there are specific examples of objects that exhibit
comparably large discrepancies \citep{nw96}. Moreover, if J0927 is a
product of a recent merger the velocity dispersion of the narrow line
region gas may not be a good tracer of the stellar velocity
dispersion.

\section{Discussion \& Conclusions}\label{S_conclude}

\subsection{The Physical Picture}\label{S_picture}

The evolution of a binary orbit is determined by the relative
efficiency of processes that can transport orbital angular momentum,
such as stellar and gas dynamical processes and in later stages, the
emission of gravitational radiation \citep{bbr}. In the binary model
considered here stellar processes are expected to be inefficient and
to operate on time scales comparable to the Hubble time
\citep{berczik06,sesana07,perets07}. A scenario in which large amounts
of cold gas are present within the sub-parsec binary orbit is not
favored in the context of current understanding of these systems
\citep{armitage02, milos05, mm06}. Feedback from the AGN and binary
torques are expected to efficiently heat and disperse the gas. As a
consequence, the gaseous dynamical friction does not affect the
orbital evolution of the binary in this phase.  The gas outside of the
SBHB orbit, in the circumbinary disk, can exert torques on the binary
and \citet{cuadra08} find that only binaries with mass $\lesssim
10^{7} \Msun$ can coalesce within a Hubble time due to this effect
\citep[see also][]{hayasaki08}, while more massive SBHBs, like the one
assumed here, evolve on longer time scales. If stellar and gas
dynamical mechanisms for angular momentum transport are inefficient,
the evolution of the sub-parsec binary orbit will be determined by the
emission of gravitational radiation and the assumption of the circular
orbit is justified. Note however that if high, near-Eddington mass
accretion rates onto the binary are plausible, interactions with the
circumbinary disk may drive the evolution of the orbital separation
and eccentricity on time scale $\sim {\rm few}\times t_{\rm
visc}\approx 10^8\,{\rm yr}$, shorter than $t_{\rm gw}$ (where $t_{\rm
visc}$ and $t_{\rm gw}$ are, respectively, the viscous time scale of
the disk at the inner edge and the gravitational wave decay
time). Realistically, $t_{\rm visc}$ will be longer because the
structure of the disk and consequently, accretion rate will be
affected by the presence of the binary. If so, the time scale given by
$t_{\rm visc}$ can be considered a conservative lower limit on the
life span of the binary and the assumption of circularity (or more
precisely, moderate eccentricity) is still a plausible one. The time
scale for orbital decay of the binary due to the emission of
gravitational waves is      
\begin{equation}
t_{\rm gw}  
\approx 1.4\times10^{10}\;{M_{2,8}\over (1+q)^5} \left({0.1\over q}\right)^2 \left({\sin i\over \sin 45^\circ}\;{\sin \phi\over \sin 45^\circ}\right)^8\;{\rm yr}\, .
\label{eq3}
\end{equation}
Since $t_{\rm gw}$ is a sensitive function of $\sin i$ and $\sin
\phi$, it may be shorter than the Hubble time if we are observing the
binary at low inclination or close to conjunction, or alternatively if
the binary orbit has eccentricity e $>$ 0.1.

\begin{figure}[t]
\epsscale{1.0} 
\plotone{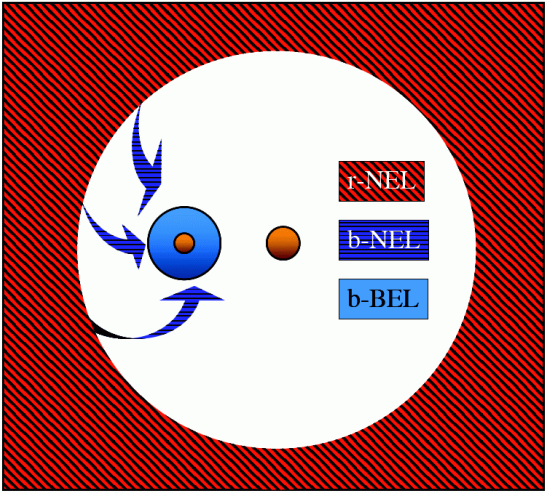}
\caption{
Illustration of the innermost region of a circumbinary disk
after the binary has cleared a low density ``hole'' in the center
(top view, not drawn to scale). In the context of the model proposed here, the
r-NELs are associated with the circumbinary disk, b-BELs with the
disk surrounding the less massive secondary SBH, and b-NELs with the
accretion streams flowing from the inner edge of the circumbinary
disk towards the disk of the secondary. Accretion occurs preferentially
on the secondary SBH, rendering the primary quiescent or much
fainter in comparison.
\label{fig_sketch}}
\end{figure}

We thus assume that the binary described here is surrounded by a
circumbinary disk, as illustrated in Figure~\ref{fig_sketch}. As the
binary orbit decays, the inner rim of the disk follows it inward until
the time scale for orbital decay by gravitational radiation becomes
shorter than the viscous time scale. In the model presented here,
$t_{\rm gw} > t_{\rm visc}$, implying that the binary has not
detached from the circumbinary disk and can still draw matter from
it. Moreover, detailed calculations have shown that for small mass
ratios (q $\lesssim {\rm 1/few}$) accretion occurs preferentially onto the
lower mass object which, as a consequence, will be more luminous and
easier to detect \citep{al96,gould00}. For example, \citet{hayasaki07}
find a significant inversion of accretion rates, $\dot{M}_2/\dot{M}_1
= 3.25$ for a $q = 0.5$ binary. This effect is consistent with the
picture proposed here that the single set of broad emission lines
observed in J0927 is associated with the secondary SBH. On the other
hand, a lower limit of $q > 0.01$ can be placed on the binary mass
ratio, given our choice $M_2 = 10^8 \Msun$ and the expected upper
limit to black hole masses, $M_{\rm max} \sim 10^{10} \Msun$
\citep{nt08}.

In addition to simulations of SBHBs, the accretion flow between a
circumbinary disk and the binary has been modeled in simulations of
disks surrounding stellar, T~Tauri binaries \citep{gunther02,
gunther04}. The dynamics of these two types of systems are, in fact,
quite similar. In particular, a common result of the above simulations
is that the accretion on individual binary members is mediated by one
or more accretion streams, implying that these may arise as a general
property of circumbinary accretion flows. The detailed structure and
spectroscopic signatures of such flows are unknown, nevertheless, we
propose the following simple picture for the system considered
here. Because the gas in accretion streams flowing towards the
secondary SBH has a nonzero angular momentum, it is expected to form a
small accretion disk surrounding the secondary SBH prior to plunging
into it. Given the uniform and monotonic orbital evolution of the
binary, this may be a long lived and stable phenomenon, where the rate
of accretion onto $M_2$ (i.e., the surface density of the disk
surrounding the black hole) will be determined by the flow rate of
matter in the gas streams. \citet{mm06} find the flow rate in such a
binary system to be of order of $10\%$ of the accretion rate onto a
single black hole with a mass equal to that of the binary. Some of the
gas in the flow is accreted by the black holes, while the rest
develops eccentric orbits, may collide with the inner rim of the
circumbinary disk and leave the binary system in form of the high
velocity outflows \citep{armitage02,mm06}. While the details of this
process remain to be modeled, we suggest in the context of our
proposed scenario that a negligibly small fraction of the gas will be
accreted by the primary black hole, given the angular momentum
``barrier'' that the gas experiences. Consequently, an AGN associated
with $M_1$ may either be faint or have short lived accretion
phases. This implies that for the binary mass ratio of $q = 0.1$, the
accretion rate onto the secondary is less than or equal to its
Eddington limit ($\dot M_2\lesssim {\dot M_{\rm Edd,\,2}}$). Thus, the
necessary conditions to establish a long term accretion process on
$M_2$ exist, though the realistic physical picture may be more complex
due to radiative feedback from the AGN.

The accretion flow within the Roche lobe of the secondary would 
resemble the accretion flow onto a single supermassive
black hole in an AGN, i.e., its optical spectroscopic signature will
be that of an AGN, exhibiting a spectrum with a blue, featureless
continuum and broad, permitted emission lines. This picture is
consistent with the properties of the observed b-BELs --- if the FWHM
of these lines are interpreted to roughly represent the size of the
emission region around the secondary, it follows that the size of 
this region is $\sim 0.1a$ (assuming $i = 45^\circ$). Using the 
\citet{eggleton83} approximation we estimate the effective Roche lobe radius
of the secondary to be $R_{L2} / a = 0.21$ for $q = 0.1$. Thus, the BLR 
is bound to the secondary SBH, consistent with the expectation that 
any gas beyond the Roche lobe of the secondary should be tidally 
truncated by the primary.

Binaries with moderate eccentricities and mass ratios that are not
extreme are expected to truncate their circumbinary disks at an inner
radius of about twice the binary semi-major axis, and hence, the
streams from the circumbinary disk to the secondary will flow over a
region of size comparable to that of the binary orbit. The stream
properties will be intermediate between the physical properties of the
secondary's BLR and the NLR of the host galaxy; \citet{dotti08}, for
example, estimate an {\it average} density in such a circumbinary
region in the range $2-8\times 10^6\,{\rm cm^{-3}}$.  Of course the
density of the gas streams themselves will be higher but they will
likely be surrounded by lower-density envelopes resulting
from expansion or ablation of the gas due to illumination and heating
by the accreting black hole. The gas in the streams will be
photoionized by the AGN continuum and the resulting emission-line
spectrum may have the following properties: $(i)$ It will consist of
permitted lines, as well as forbidden lines from the lower-density
parts of the flow, such as the ionized skin of the
streams.  Due to proximity in velocity space, the lines from the
stream should be shifted to approximately the velocity of the broad
emission lines from the vicinity of the secondary.  $(ii)$ The
profiles of the lines from this stream will be narrower than the
broad, permitted lines and asymmetric, since they trace the {\it
  emissivity weighted} distribution of the spatially confined stream
of gas\footnote{The asymmetry may arise since in general case the
  stream geometry and velocity will not be symmetric with respect to
  the observer.  While the exact profile shapes of these lines will
  depend on the photoionization effects (not considered here), these
  are not expected to give rise to symmetric line profiles, in
  general.}.  $(iii)$ The line shifts, and perhaps also the
asymmetries will be variable over the orbital cycle of the binary.  We
propose that the b-NELs, with their variety of shifts and widths,
originate in this part of the flow. Circumstantial evidence in support
of the proposed picture is also provided by the relative intensities
of the b-NEL and r-NEL lines. By comparing the relative line
intensities reported by \citet{komossa08} with the photoionization
models of \citet{nagao01,nagao02} we find that the observed
[\ion{Ne}{3}]/[\ion{O}{2}] ratios suggest a higher density (by 1--2
orders of magnitude) in the b-NLR compared to the r-NLR. Moreover, the
[\ion{O}{3}]/H$\beta$ ratios suggest a higher ionization parameter in
the former region by a factor of several, which is also reflected in
the [\ion{Ne}{3}]/[\ion{Ne}{5}] ratio. These differences in density 
and ionization are consistent with our hypothesis that the r-NLR 
is associated with gas in the nucleus of the host galaxy while 
the b-NLR is associated with denser gas that is part
of the accretion flow onto the secondary black hole. The above conclusions are
subject to some uncertainty because the results of photoionization
models are sensitive to a variety of input parameters, such as the
spectral energy distribution of the ionizing continuum. 

\subsection{The Recoiling Black Hole Scenario Revisited}\label{S_revisit}

Let us consider the characteristic time scales relevant for the
recoiling SBH model with the parameters adopted by
\citet{komossa08}. The time during which a recoiling SBH of mass $\sim
6 \times 10^8\,\Msun$ appears as an AGN can in principle be close to
$10^9$~yr, if the mass of the accretion disk carried along by the SBH
is comparable to its own mass \citep{loeb07}. This cannot be the case
in J0927, if its SBH charges away from the host galaxy at nearly the
maximum speed predicted by numerical relativity. To accommodate a high
velocity kick, the mass of the disk should be small enough not to slow
down the SBH, implying that $t_{\rm AGN} \ll 10^9$~yr. Indeed,
\citet{bl08} calculate that the mass of the gas disk carried by the
recoiling SBH in J0927 is $\sim 2 \%$ of the SBH mass, which would be
accreted in only $\sim 10^7$~yr, if the accretion rate is 0.1 of the
Eddington limit.

If the FWHM of the [\ion{Ne}{3}] line is indicative of the expansion
of the accretion disk due to outward transport of angular momentum, as
suggested in the context of the recoiling SBH model, then the final
disk radius is $\sim$7 times larger than that immediately after the
recoil. This factor is based on a comparison of the FWHM of the
[\ion{Ne}{3}] line \citep[reported by][]{komossa08} and the line of
sight recoil velocity\footnote{Assuming that the recoil velocity and
  the FWHM of the [\ion{Ne}{3}] line indicate the truncation radius of
  the disk before and after the expansion. Larger factors are expected
  if the radius of marginal self-gravity is considered instead to
  obtain the former value.}.  A disk expansion should be accompanied
by a drop in surface density and an accretion rate reduction by a
factor of $> 7^2$ (assuming the same disk scale height before and
after the expansion). Even if the initial luminosity of this system
was close to the Eddington limit, it is now a factor of 100 lower,
implying a minimum SBH mass of $\sim5\times10^9\,\Msun$, given the
estimated bolometric luminosity. It also follows from conservation of
angular momentum that in process of disk expansion a sizable portion
of the disk mass will be accreted, $M_{\rm acc} \sim M_{\rm disk} /
\sqrt{7}$. Given the SBH recoil velocity and the estimated accretion
time scale of $\sim 10^7$~yr \citep{bl08}, this implies that, by the
time this fraction of disk mass is accreted, the receding AGN should
be at least few tens of kiloparsecs away from the host galaxy. This
poses a challenge for the recoiling SBH model given the requirement
for the AGN to power the observed emission lines in both its own BLR
and NLR and the NLR bound to the host galaxy. More specifically, the
ionization parameter of the r-NLR should be considerably lower than
what is observed (see our comparison with photoionization models in
\S\,\ref{S_picture}). We note that origin of the widths and shifts of
the [NeIII] and other b-NELs in an outflow or the swept up ISM gas 
cannot be excluded. In such a scenario the geometry and
velocity distribution of the line-emitting gas is likely to be very
complex.

A detection of an AGN recoiling at nearly the maximum speed predicted
by numerical relativity requires a combination of coincidences: (a)
its recoil velocity vector must lie close to our line of sight, (b)
the parameters of the pre-coalescence binary should have been such as
to give a recoil speed close to the maximum, and (c) we should be
observing the object in a relatively narrow time window after the
recoil. This would also imply a larger population of systems at lower
recoil speeds, which were not found by \citet{bonning07}.  Indeed,
\citet{dotti08} calculate that SBHBs with parameters similar to those
considered here can be $\sim$100 times intrinsically more common than
the recoiling SBHs. A related question, then, is why have there been
no more discoveries of SBHBs in the SDSS. The observational biases in
the case of subparsec SBHBs are not understood due to uncertainties in
the structure and properties of their nuclear regions. The Doppler
shift signature may not be observable in all subparsec binaries.  
The majority of them may either be quiescent or they may
exhibit a different signature that we still do not recognize, thus,
making the expected (a posteriori) discovery rate for this class of
objects difficult to evaluate.

\subsection{Possible Observational Tests}\label{S_tests}

The observational property of J0927 that presents the biggest
challenge to any model are its two sets of narrow emission-lines. It
seems that by coincidence the [\ion{O}{3}]~$\lambda$4959 line at
$z=0.713$ overlaps with the [\ion{O}{3}]~$\lambda$5007 line at
$z=0.698$. Moreover, the low signal-to-noise ratio (S/N) at the blue
end of the original SDSS spectrum makes the line profiles difficult to
characterize. New optical spectra with a higher S/N and spectral
resolution are needed to characterize the line profiles and make
better measurements of some line strengths (especially
[\ion{Ne}{5}]). Complementary near-IR spectra in the J-band could show
the profile of the H$\alpha$ line and additional narrow, forbidden
lines of important diagnostic value.

Spectroscopic monitoring on time scales of years to decades could
provide the most direct test of the binary hypothesis.  According to
the SBHB model, the emission lines associated with the BLR of the
secondary SBH (b-BELs) will shift at a rate of
%
%
$$
{du_2\over dt} \approx 88\; {(1+q)^2\over M_{2,8}} \left({q\over 0.1}\right) 
\left({\sin i\over \sin 45^\circ}\right)^{-3} 
\hbox to 4 em{\hss}
$$
\begin{equation}
\hbox to 4 em{\hss} \times
\left({\sin \phi\over \sin 45^\circ}\right)^{-4} \left({\cos \phi\over \cos 45^
\circ}\right)\;{\rm km~s^{-1}~yr^{-1}}\, .
\label{eq_shift}
\end{equation}
Thus, at $\phi=45^{\circ}$ the b-NELs would shift by $\sim 100~{\rm
km~s}^{-1}$ in about a year\footnote{However, note that there is a
range of non-extreme parameter values for which the normalization
factor in equation~\ref{eq_shift} takes a lower value and
consequently, the monitoring period may need to be
longer.}. Spectroscopic and imaging observations at high angular
resolution with the {\it Hubble Space Telescope} the ({\it HST}) are
necessary in order to measure the redshift of the host galaxy,
determine whether its morphology shows signs of a recent merger, and
provide a check against the possibility of projection of two otherwise
unrelated AGNs along the line of sight. Also, in case of the recoiling
SBH, {\it HST} may resolve an off center AGN, depending on the
orientation of the recoil velocity vector with respect to the line of
sight. In the case of a binary scenario, if the two black holes (or at
least the secondary) are radio emitters, their orbital motion may be
detected with the $VLBA$. At the redshift of J0927, the orbital
separation of the binary with parameters scaled as in eq~(\ref{eq1}),
translates to an angular separation on the sky of about $\sim
20\,\mu{\rm as}$. This is comparable to the astrometric precision
achieved with the VLBA and for a higher mass binary may approach the
expected spatial resolution of the {\it Square Kilometre Array}.

The confirmation of either model would be a major step towards our
understanding of subparsec binaries or recoiling SBHs. A binary holds
great promise for revealing the physics of gas in the nuclear region
and the accretion signatures in the pre-coalescence phase of its
evolution. Discovery of a recoiling SBH, on the other hand, has
important implications for understanding the demographics of SBHs and
their cosmological spin evolution. Both scenarios have direct
ramifications for the rate of SBHB coalescences and the exploration of
the binary parameter space, of large importance for gravitational wave
observatories such as the {\it Laser Interferometer Space Antenna}.

\acknowledgments

We thank Cole Miller for useful discussions, and Chris Reynolds and 
Sean O'Neill for helpful comments. TB thanks the UMCP-Astronomy 
Center for Theory and Computation Prize Fellowship program for support.



\begin{thebibliography}{99}


\bibitem[\protect\citeauthoryear{Adelman-McCarthy et al.}{2007}]{am07} 
Adelman-McCarthy, J.~K., et al.\ 2007, \apjs, 172, 634 

\bibitem[\protect\citeauthoryear{Armitage \& Natarajan}{2002}]{armitage02} Armitage,
P.~J., \& Natarajan, P.\ 2002, \apjl, 567, L9


\bibitem[\protect\citeauthoryear{Artymowicz \& Lubow}{1996}]{al96} Artymowicz, P., \& Lubow, S.~H.\ 1996, \apjl, 467, L77 

\bibitem[\protect\citeauthoryear{Baker et al.}{2008}]{baker08} Baker, J.~G., Boggs, 
W.~D., Centrella, J., Kelly, B.~J., McWilliams, S.~T., Miller, M.~C., 
\& van Meter, J.~R.\ 2008, \apjl, 682, L29 

\bibitem[\protect\citeauthoryear{Baker et al.}{2007}]{baker07} Baker, J.~G., Boggs, 
W.~D., Centrella, J., Kelly, B.~J., McWilliams, S.~T., Miller, M.~C., \& van Meter, J.~R.\ 2007, \apj, 668, 1140 

\bibitem[\protect\citeauthoryear{Baker et al.}{2006}]{baker06} Baker, J.~G., Centrella, 
J., Choi, D.-I., Koppitz, M., van Meter, J.~R., \& Miller, M.~C.\ 2006, \apjl, 653, L93 

\bibitem[\protect\citeauthoryear{Begelman, Blandford, \& Rees}{Begelman et al.}{1980}]{bbr}
Begelman, M.~C., Blandford, R.~D., \& Rees, M.~J.\ 1980, \nat, 287, 307

\bibitem[\protect\citeauthoryear{Berczik et al.}{2006}]{berczik06} Berczik, P., Merritt, D., Spurzem, R. \& Bischof, H.-P., 2006, 
\apjl, 642, 21

\bibitem[\protect\citeauthoryear{Blecha \& Loeb}{2008}]{bl08} Blecha, L., \& Loeb, A.\ 2008, \mnras, 390, 1311  


\bibitem[\protect\citeauthoryear{Bogdanovi{\'c} et al.}{2007}]{bogdanovic07} Bogdanovi{\'c}, 
T., Reynolds, C.~S., \& Miller, M.~C.\ 2007, \apjl, 661, L147



\bibitem[\protect\citeauthoryear{Bonning et al.}{2007}]{bonning07} Bonning, E.~W., 
Shields, G.~A., \& Salviander, S.\ 2007, \apjl, 666, L13 

\bibitem[\protect\citeauthoryear{Campanelli et al.}{2007a}]{campanelli07a} Campanelli, M., 
Lousto, C., Zlochower, Y., \& Merritt, D.\ 2007a, \apjl, 659, L5 

\bibitem[\protect\citeauthoryear{Campanelli et al.}{2007b}]{campanelli07b} Campanelli, M., 
Lousto, C., Zlochower, Y., \& Merritt, D.\ 2007b, Physical Review Letters, 98, 231102 

\bibitem[\protect\citeauthoryear{Colpi et al.}{2007}]{colpi07} Colpi,
M., Dotti, M., Mayer, L., \& Kazantzidis, S.\ 2007, in "2007 STScI
Spring Symposium: Black Holes", eds. M. Livio \& A. M. Koekemoer,
Cambridge University Press (arXiv:0710.5207)

\bibitem[\protect\citeauthoryear{Cuadra et al.}{2008}]{cuadra08} Cuadra, J., Armitage, 
P.~J., Alexander, R.~D., \& Begelman, M.~C.\ 2008 (arXiv:0809.0311) 
 
 \bibitem[\protect\citeauthoryear{Devecchi et al.}{2008}]{devecchi08} Devecchi, B., Rasia, 
E., Dotti, M., Volonteri, M., \& Colpi, M.\ 2008 (arXiv:0805.2609) 



\bibitem[\protect\citeauthoryear{Dotti et al.}{2008}]{dotti08} Dotti,
M., Montuori, C., Decarli, R., Volonteri, M., Colpi, M., \& Haardt,
F.\ 2008 (arXiv:0809.3446)

\bibitem[\protect\citeauthoryear{Eggleton}{1983}]{eggleton83} Eggleton, P.~P.\ 1983, \apj, 
268, 368 

\bibitem[\protect\citeauthoryear{Elvis et al.}{1994}]{elvis94}
Elvis, M., Wilkes, B. J., McDowell, J. C., Green, R. F., Bechtold, J.,
Willner, S. P., Oey, M. S., Polomski, E., \& Cutri, R. 1994, \apjs, 95, 1




\bibitem[\protect\citeauthoryear{Ferrarese \& Ford}{2005}]{ff05} Ferrarese, L., \& Ford, H.\ 2005, Space Science Reviews, 116, 523 



\bibitem[\protect\citeauthoryear{Gonz{\'a}lez et al.}{2007}]{gonzalez07b} Gonz{\'a}lez, 
J.~A., Sperhake, U., Br{\"u}gmann, B., Hannam, M., \& Husa, S.\ 2007, Physical Review Letters, 98, 091101


\bibitem[\protect\citeauthoryear{Gould \& Rix}{2000}]{gould00} Gould, A., \& Rix, H.\
2000, \apjl, 532, L29

\bibitem[\protect\citeauthoryear{Gualandris \& Merritt}{2008}]{gm08} Gualandris, A., \& Merritt, D.\ 2008, \apj, 678, 780 

\bibitem[\protect\citeauthoryear{G{\"u}nther \& Kley}{2002}]{gunther02} G{\"u}nther, R., \& Kley, W.\ 2002, \aap, 387, 550 

\bibitem[\protect\citeauthoryear{G{\"u}nther et al.}{2004}]{gunther04} G{\"u}nther, R., Sch{\"a}fer, C., \& Kley, W.\ 2004, \aap, 423, 559 


\bibitem[\protect\citeauthoryear{Hayasaki}{2008}]{hayasaki08} Hayasaki, K.\ 2008 (arXiv:0805.3408) 

\bibitem[\protect\citeauthoryear{Hayasaki et al.}{2007}]{hayasaki07} Hayasaki, K., 
Mineshige, S., \& Sudou, H.\ 2007, \pasj, 59, 427 

\bibitem[\protect\citeauthoryear{Herrmann et al.}{2007a}]{herrmann07a} Herrmann, F., Hinder, 
I., Shoemaker, D., \& Laguna, P.\ 2007a, Classical and Quantum Gravity, 24, 33 

\bibitem[\protect\citeauthoryear{Herrmann et al.}{2007b}]{herrmann07b} Herrmann, F., Hinder, 
I., Shoemaker, D., Laguna, P., \& Matzner, R.~A.\ 2007b, \apj, 661, 430 
 

\bibitem[\protect\citeauthoryear{Kaspi et al.}{2005}]{kaspi05} Kaspi, S., Maoz, D., Netzer, H., Peterson, B.~M., Vestergaard, M., 
\& Jannuzi, B.~T.\ 2005, \apj, 629, 61 


\bibitem[\protect\citeauthoryear{Kocsis \& Loeb}{2008}]{kl08} Kocsis, B., \& Loeb, A.\ 2008, Physical Review Letters, 101, 041101 

\bibitem[\protect\citeauthoryear{KZL08}{}]{komossa08} Komossa, S., Zhou, H., 
\& Lu, H.\ 2008, \apjl, 678, L81 (KZL08)

\bibitem[\protect\citeauthoryear{Koppitz et al.}{2007}]{koppitz07} Koppitz, M., Pollney, 
D., Reisswig, C., Rezzolla, L., Thornburg, J., Diener, P., 
\& Schnetter, E.\ 2007, Physical Review Letters, 99, 041102

\bibitem[\protect\citeauthoryear{Lippai et al.}{2008}]{lippai08} Lippai, Z., Frei, Z., 
\& Haiman, Z.\ 2008, \apjl, 676, L5 

\bibitem[\protect\citeauthoryear{Loeb}{2007}]{loeb07} Loeb, A.\ 2007, Physical Review 
Letters, 99, 041103 

\bibitem[\protect\citeauthoryear{MacFadyen \&
Milosavljevi\'c}{2008}]{mm06} MacFadyen, A.~I., \& Milosavljevi{\'c},
M.\ 2008, \apj, 672, 83


\bibitem[\protect\citeauthoryear{Merritt et al.}{2004}]{merritt04} Merritt, D., 
Milosavljevi{\'c}, M., Favata, M., Hughes, S.~A., \& Holz, D.~E.\ 2004, \apjl, 607, L9 

\bibitem[\protect\citeauthoryear{Milosavljevi\'c \& Phinney}{2005}]{milos05} Milosavljevi\'c, M., \& Phinney, E.S., 2005, \apjl, 622, 93


\bibitem[\protect\citeauthoryear{Mohayaee et al.}{2008}]{moh08} Mohayaee, R., Colin, 
J., \& Silk, J.\ 2008, \apjl, 674, L21

\bibitem[\protect\citeauthoryear{Nagao et al.}{2002}]{nagao02} Nagao, T., Murayama, T., Shioya, Y., \& Taniguchi, Y. 2002, \apj, 567, 73

\bibitem[\protect\citeauthoryear{Nagao et al.}{2001}]{nagao01} Nagao, T., Murayama, T., \& Taniguchi, Y. 2001, \apj, 546, 744

\bibitem[\protect\citeauthoryear{Natarajan \& Treister}{2008}]{nt08} Natarajan, P., \& Treister, E.\ 2008 (arXiv:0808.2813) 

\bibitem[\protect\citeauthoryear{Nelson}{2000}]{nelson00} Nelson, C.~H.\ 2000, \apjl, 544, L9

\bibitem[\protect\citeauthoryear{Nelson \& Whittle}{1996}]{nw96} Nelson, C.~H., \& Whittle, M.\ 1996, \apj, 465, 96 

\bibitem[Perets et al.(2007)]{perets07} Perets, H.~B., Hopman, C., \&
Alexander, T.\ 2007, \apj, 656, 709


\bibitem[\protect\citeauthoryear{Redmount \& Rees}{1989}]{rr89} Redmount, I.~H., \& Rees, M.~J.\ 1989, Comments on Astrophysics, 14, 165

\bibitem[\protect\citeauthoryear{Rezzolla et al.}{2008}]{rezzolla08} Rezzolla, L., Dorband, E.~N., Reisswig, C., Diener, P., Pollney, D., Schnetter, E., \& Szil{\'a}gyi, B.\ 2008, \apj, 679, 1422 

\bibitem[\protect\citeauthoryear{Schnittman \& Buonanno}{2007}]{sb07} Schnittman, J.~D., \& Buonanno, A.\ 2007, \apjl, 662, L63 

\bibitem[\protect\citeauthoryear{Schnittman \& Krolik}{2008}]{sk08} Schnittman, J.~D., \& Krolik, J.~H.\ 2008, \apj, 684, 835 

\bibitem[\protect\citeauthoryear{Sesana et al.}{2007}]{sesana07} Sesana, A., Haardt, F., \& Madau, P.\ 2007, \apj, 660, 546


\bibitem[\protect\citeauthoryear{Shields \& Bonning}{2008}]{sb08}  Shields, G.~A., \& Bonning, E.~W.\ 2008, \apj, 682, 758 

\bibitem[\protect\citeauthoryear{Tremaine et al.}{2002}]{tremaine02} Tremaine, S., et al.\ 2002, \apj, 574, 740 


\end{thebibliography}
\end{document}